\documentclass[doublespacing]{elsart}

\usepackage{graphicx}

\usepackage{amssymb,amsmath}

\journal{Physics Letters A}

\newcommand{\ie}{i.e.}

\begin{document}

\begin{frontmatter}


\title{A comparison between matter wave and light wave interferometers for the detection of gravitational waves}


\author[ERGA]{P. Delva\corauthref{pac}},
\corauth[pac]{Corresponding author.}
\ead{pacome.delva@obspm.fr}
\author[ERGA]{M.-C. Angonin} and
\ead{m-c.angonin@obspm.fr}
\author[ERGA]{P. Tourrenc}
\ead{pht@ccr.jussieu.fr}

\address[ERGA]{Universit\'e P. et M. Curie, ERGA, case 142, 4, place Jussieu, F-75252, Paris Cedex 05, France}

\begin{abstract}
We calculate and compare the response of light wave interferometers and matter wave interferometers to gravitational waves. We find that metric matter wave interferometers will not challenge kilometric light wave interferometers such as Virgo or LIGO, but could be a good candidate for the detection of very low frequency gravitational waves.
\end{abstract}

\begin{keyword}
Gravitational waves detection \sep Matter waves \sep Interferometry
\PACS 04.30.-w \sep 04.80.Nn \sep 95.55.Ym \sep 03.75.-b \sep 39.20.+q
\end{keyword}
\end{frontmatter}


\section{Introduction}

The extreme sensitivities required to detect gravitational waves still represent a challenge. LIGO interferometers\footnote{http://www.ligo.caltech.edu/} have reached their nominal sensitivity and have begun scientific runs giving upper limits of gravitational waves in various cases : gravitational bursts, stochastic background and periodic waves \cite{abbott05b,abbott05a,abbott04}. Virgo\footnote{http://www.ego-gw.it/} is now in the commissioning phase; its sensitivity should be better than LIGO's at low frequencies (around 10 Hz).

However despite important technological improvements in many domains, the present detectors will not be able to observe the very low frequency sources of astrophysical interest. The LISA project \cite{lisa00}, a space-based laser interferometer with 5 millions kilometers arms, is presently the best known solution to improve the detectors capabilities at very low frequencies. The challenge is impressive, we thus believe that new technologies will be necessary\ for further developments in the future.

In 2004, Chiao and Speliotopoulos \cite{chiao04a} proposed to use matter-wave interferometers as gravitational wave detectors (the so-called MIGO, Matter-wave Interferometric Gravitational-wave Observatory). They claimed that MIGO could reach the same sensitivity as LISA or LIGO in a much more compact configuration. Their results are discussed in papers from Roura et al. \cite{roura04} and Foffa et al. \cite{foffa06}. The various authors find divergent results. These differences originate from different physical interpretations of the various coordinate systems which were used and from the different boundary conditions which were assumed.

In the present paper we consider various matter-wave interferometers. For each of them, we estimate the\ order of magnitude of the phase shift due to a periodic gravitational wave. We compare the sensitivity of matter wave interferometers (MWI) and light wave interferometers (LWI). Several boundary conditions are considered~: fixed or free mirrors. In each case we clearly point forward the corresponding comoving coordinates. We emphasize the role of the three mostly relevant parameters, \ie{} the atomic flux, the atomic speed and the number of atomic bounces in the arms of the interferometers when similar to Fabry-Perot cavities in optics.

In the present paper, we do not consider the technical feasibility. Our goal is just to compare MWIs and LWIs in order to become familiar with MWIs and to discover the conditions necessary for matter wave interferometers to be useful. Once these conditions are put forward it might be possible to have an opinion on the future feasibility of such detectors. For this next step, an expertise in high-tech experiments, cold atoms and matter wave interferometry is crucial, as well as a high dose of optimism.

Now, before we start, let us recall that a gravitational wave is a perturbation, $h_{\mu \nu}$, in the local metric~:

\begin{equation}
\label{ORDRE1}
g_{\mu \nu} = \eta_{\mu \nu} + h_{\mu \nu}
\end{equation}

where $\eta _{\mu \nu }$ is the usual Minkowski metric and the greek indices run from 0 to 3. In the sequel  $h$ refers to the order of magnitude of the biggest $\left\vert h_{\mu \nu} \right \vert$ (\ie{} $\left\vert h_{\mu \nu }\right\vert \lesssim h<<1)$.

\section{The quantum phase}

In this section we consider the matter wave associated to a particle of mass $m$  (for a light wave $m=0$). We investigate the perturbation of the quantum phase due to the perturbation $h_{\mu \nu}$ of the metric.

First we consider the propagation of the wave between two mirrors or two beam splitters A and B. The quantum phase $\phi $ is developed up to the first order relatively to $h$~: $\phi = \phi_o + \delta \phi$ where $\phi_o$ is the unperturbed quantum phase while $\delta \phi$ is the perturbation of order $h$. The phase $\phi$ is assumed to be a solution of the eikonal equation~: $g^{\mu \nu} \phi_{,\mu} \phi_{,\nu} = (m c / \hbar)^2$, where $\hbar$ is Planck's constant and $c$ the light velocity in vacuum. Therefore the unperturbed solution is $\phi_o = k_\mu x^\mu + cst$ with $k_\mu = \partial_\mu \phi_o = \eta_{\mu \nu} p^\nu / \hbar$, where $p^\nu$ is the unperturbed particle momentum. For a matter wave $p^\nu = \gamma m c \left( 1,\vec{v} / c \right)$, where $\vec{v}$ is the group velocity of the wave and $\gamma = \dfrac{1}{\sqrt{1 - \vec{v}^2 / c^2}}$. Let us notice that the atom energy is $E = \hbar c k^0$.

\begin{figure}[h]
\begin{center}
\includegraphics[width=0.5\linewidth]{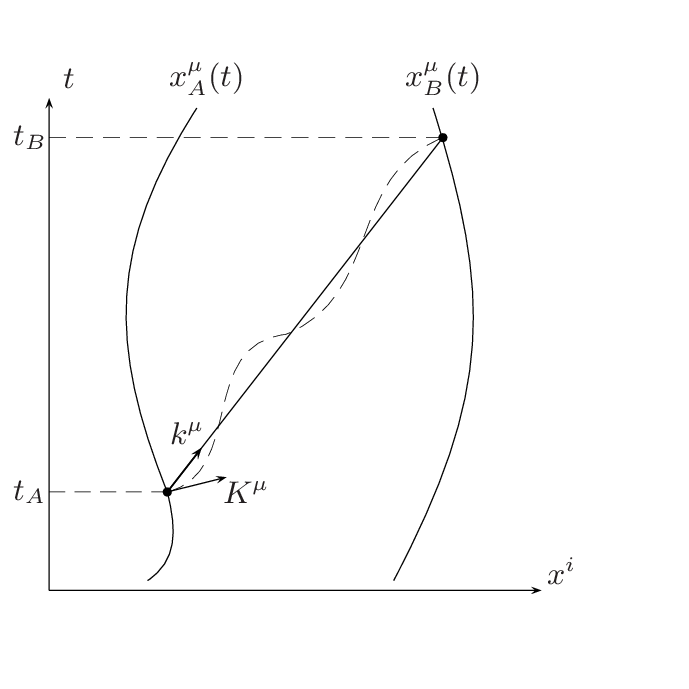}
\caption{The matter wave propagates between the two optical elements A and B (dashed line). The unperturbed trajectory is the straight line.}
\label{ET}
\end{center}
\end{figure}

The wordlines of the mirrors or the beam splitters A and B are perturbed by the GW. Let call $x^\mu_A = x^\mu_A (t)$ and $x^\mu_B = x^\mu_B (t)$ their trajectories. Knowing $k^\mu$ and $t_B$ (the arrival time of the atom, we can deduce $t_A$ (its departure time)~: $x_B^\mu - x_A^\mu = \alpha k^\mu$, where $\alpha$ is a constant. This is illustrated on Fig.~\ref{ET}. We define $\left[ \phi \right]_{A}^{B} \equiv \phi \left[ x_B^\mu \left( t_B \right) \right] - \phi \left[ x_A^\mu \left( t_A \right) \right]$. In order to calculate $\left[ \phi \right]_A^B$ we use the method described in \cite{linet76} and \cite{stodolsky79}. The solution reads~:

\begin{equation}
\label{PHITOT}
\left[ \phi \right]_{A}^{B} = \left[ \phi _o \right]_A^B + \left[ \delta \phi \right]_A^B
\end{equation}

where~:

\begin{equation}
\label{PHI0}
\left[ \phi _{o}\right] _{A}^{B}=k_{\mu }x_{B}^{\mu }-k_{\mu }x_{A}^{\mu}
\end{equation}

\begin{equation}
\label{PHI}
\left[ \delta \phi \right] _{A}^{B}=\frac{\hbar c^{2}}{2} \int_{t_{A}}^{t_{B}}h_{\mu \nu }k^{\mu }k^{\nu }\frac{\d t}{E}
\end{equation}

The integration in (\ref{PHI}) is performed along the unperturbed trajectory (the straight line of Fig.~\ref{ET}).

The phase is a scalar, therefore $\left[ \phi \right]_A^B$ is independent of the coordinate system, although the decomposition~(\ref{PHITOT}) is not. A first order coordinate transformation results in a first order change of the functions $x^\mu_{A,B}(t)$ and a first order change of $h_{\mu \nu}$ in (\ref{ORDRE1}). Both effects have to be into account. Then, contrary to the claim of Chiao \& Speliotopoulos in \cite{chiao04a}, one can show explicitely that the result is invariant \cite{roura04}. In the sequel, for the sake of simplicity, we will choose a coordinate system where the optical elements are at rest, \ie{} $\vec{x}_A$ and $\vec{x}_B$ are independant of the time coordinate $x^0$.

\section{The free interferometer}

\subsection{The Einstein Coordinates (EC)}

\label{EC}

In the weak-field approximation, in empty space, the Einstein equations are~:

\begin{equation}
\label{EE}
R_{\mu \nu} = \frac{1}{2} \eta ^{\sigma \tau } \left( h_{\sigma \tau, \mu \nu} + h_{\mu \nu,\sigma \tau} - h_{\sigma \nu, \mu \tau} - h_{\mu \tau, \sigma \nu} \right) = 0
\end{equation}

In harmonic coordinates the condition $\partial_\nu h_\mu^\nu = \frac{1}{2} \eta_\mu^\nu \partial_\nu h_\alpha^\alpha $ holds true. The Einstein equations~(\ref{EE}) read~:

\begin{equation}
\label{EP}
\square h_{\mu \nu} = 0
\end{equation}

Gravitational waves (GW) are solutions of such a propagation equation. Far away from the sources, plane waves are solutions of (\ref{EP}). Some more constraints on $h_{\mu \nu}$ (\ie{} $\partial_{i}h_{j}^{i}=0$ and $h_{i}^{i}=0$, where latin indices run from 1 to 3) define an unique coordinate system~: we will call it the Einstein Coordinates (EC). To simplify the problem, we assume that the GW is propagating along the $z\equiv x^{3}$ axis. The only non zero components of $h_{\mu \nu }$ are $h_{rs}=h_{rs}(z-ct)$, with $r,s = 1$ or 2, then~:

\begin{equation}
\label{MTT}
\d s^2 = \eta_{\mu \nu} \d x^\mu \d x^\nu + h_{rs} \d x^r \d x^s
\end{equation}

In the slow motion limit, the space coordinates $x^{i}$ of a point mass particle satisfy the equations~:

\begin{equation}
\frac{\d^2 x^i}{\d t^2} = - \dot{h}_j^i \frac{\d x^j}{\d t} + O \left( h^2,v^2 h \right) \text{ with }\left\vert v^{i} \right\vert = \left\vert \dfrac{\d x^i}{\d t}\right\vert \ll c
\end{equation}

where $\dot{()} = \dfrac{\d}{\d t}()$.

It is easy to see that a point mass initially at rest remains at rest in the EC during the passage of a GW.

\subsection{The Michelson-Morley interferometer}

\label{PD}

We study a Michelson-Morley configuration with the arms along the $x^{1}$ and $x^{2}$ axis. The optical elements are supposed to be free of any constraints (free falling in LISA or fixed on super attenuators in Virgo). Their spatial EC remain constant with time.

\begin{figure}[h]
\begin{center}
\includegraphics[width=0.3\linewidth]{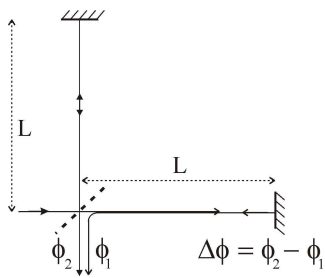}
\caption{Michelson interferometer.}
\label{MICHPIC}
\end{center}
\end{figure}

We assume\ that $h_{rs}$ is a sinusoidal function where $\Omega $ is the angular frequency of the gravitational wave. Moreover we assume that the size of the experimental device along the $z$ axis is much smaller than the gravitational wavelength $\Lambda =2\pi c/\Omega $. These conditions results in $h_{11} \left( z - ct \right) = - h_{22} \left( z - ct \right) = h_{+} \cdot \sin{\left( \Omega t\right) }$ and $h_{12} \left( z - ct \right) = h_{21} \left( z - ct \right) = h_{\times} \cdot \sin{\left(\Omega t + \varphi _{\times } \right)}$ where $h_{+}$, $h_{\times}$ and $\varphi_{\times }$ are three constants. Then, using the notations of Fig.~\ref{MICHPIC} and eqs.~(\ref{PHITOT}) and (\ref{MTT}), the phase difference between the two arms of the interferometer is~:

\begin{equation}
\label{FREE2}
\Delta \phi =-4\pi h_{+}\cdot \dfrac{V}{\Omega \lambda }\cdot \sin {\dfrac{\Omega L}{V}}\cdot \sin {\Omega t}
\end{equation}

where $V=c$ for a light wave and $V=v_{0}$, the initial group velocity, for a matter wave. $\lambda$ is the wavelength; for a matter wave, it is the De Broglie wavelength~: $\lambda = 2 \pi \hbar / m \gamma v_{0}$.

When $\Omega L/V\ll 2\pi $ the amplitude of $\Delta \phi$ in (\ref{FREE2}) is~:

\begin{equation}
\label{DIFF}
\widetilde{\Delta \phi }=4\pi | h_{+} | \cdot \frac{L}{\lambda}
\end{equation}

This formula is well known for LWIs such as Virgo or LISA \cite[p.~54]{barone00}. It holds true for a MWI. However, for a LWI, the condition $\Omega L / V \ll 2 \pi$ corresponds to $L \ll \Lambda$ while it implies $L \ll \dfrac{v_{0}}{c} \cdot \Lambda$ for a MWI.

\subsection{Light Wave Interferometers versus Matter Wave Interferometers}

\label{VS}

In Virgo or LISA, the laser source is a Nd:YAG infrared laser with $\lambda \simeq 10^{-6}$~m. Decreasing the laser wavelength in order to increase the amplitude $\widetilde{\Delta \phi}$ of the phase difference in (\ref{DIFF}) raises technical difficulties (laser stability, mirrors quality). On the other hand it seems easier to achieve in a MWI~: for instance in \cite{keith91}, the wavelength is $16$~pm. As a consequence, the phase difference in a MWI can be much larger than in a LWI, and the sensitivity appears to be much better. Unfortunately, this conclusion does not hold anymore when one considers the fundamental limit.

The signal is ultimately limited by the shot noise. The minimum phase difference that can be detected is~:

\begin{equation}
\label{SHOT}
\widetilde{\Delta \phi }\sim \frac{1}{2\sqrt{\dot{N}t}}
\end{equation}

where $\dot{N}$ is the particle flux and $t$ the integration time.

The atom flux in the experiment described in \cite{keith91} is only 70~s$^{-1}$ which is very low but in reference \cite{gustavson00} the atom flux is $10^{11}$~s$^{-1}$ for a velocity of 290~m.s$^{-1}$ (\ie{} $\lambda \simeq 10$~pm). Such a flux remains however very small compared to the photon flux in Virgo which is of order 10$^{23}$~s$^{-1}$ \cite{punturo04}. Therefore one can show that metric MWIs have to be considerably improved to compete with kilometric LWIs.

However, the LISA effective flux can be estimated of order $10^{8}$~s$^{-1}$ \cite[p. 60]{lisa00}. It is much lower than the Virgo effective flux, and even lower than the better MWI fluxes. Therefore, it seems easier to achieve with a MWI the sensitivity of LISA than the sensitivity of Virgo.

Assuming that the detection is only limited by the shot noise and that the integration times are the same, one obtains the conditions for similar
sensitivities of MWIs and LWIs from formulas (\ref{DIFF}) and (\ref{SHOT})~:

\begin{equation}
\label{COMP}
\gamma v_{0} L_{\mathrm{mw}} \sim \dfrac{ 2 \pi \hbar L_{\mathrm{lw}} }{ m \lambda_{\mathrm{lw}} } \sqrt{ \dfrac{ \dot{N}_{\mathrm{lw}} }{ \dot{N}_{\mathrm{mw}} } }
\end{equation}

where the subscripts $_{\mathrm{mw}}$ and $_{\mathrm{lw}}$ denote respectively the characteristics of the MWI and the LWI.

\begin{figure}[h]
   \begin{minipage}[l]{.47\linewidth}
      \includegraphics[width=\linewidth]{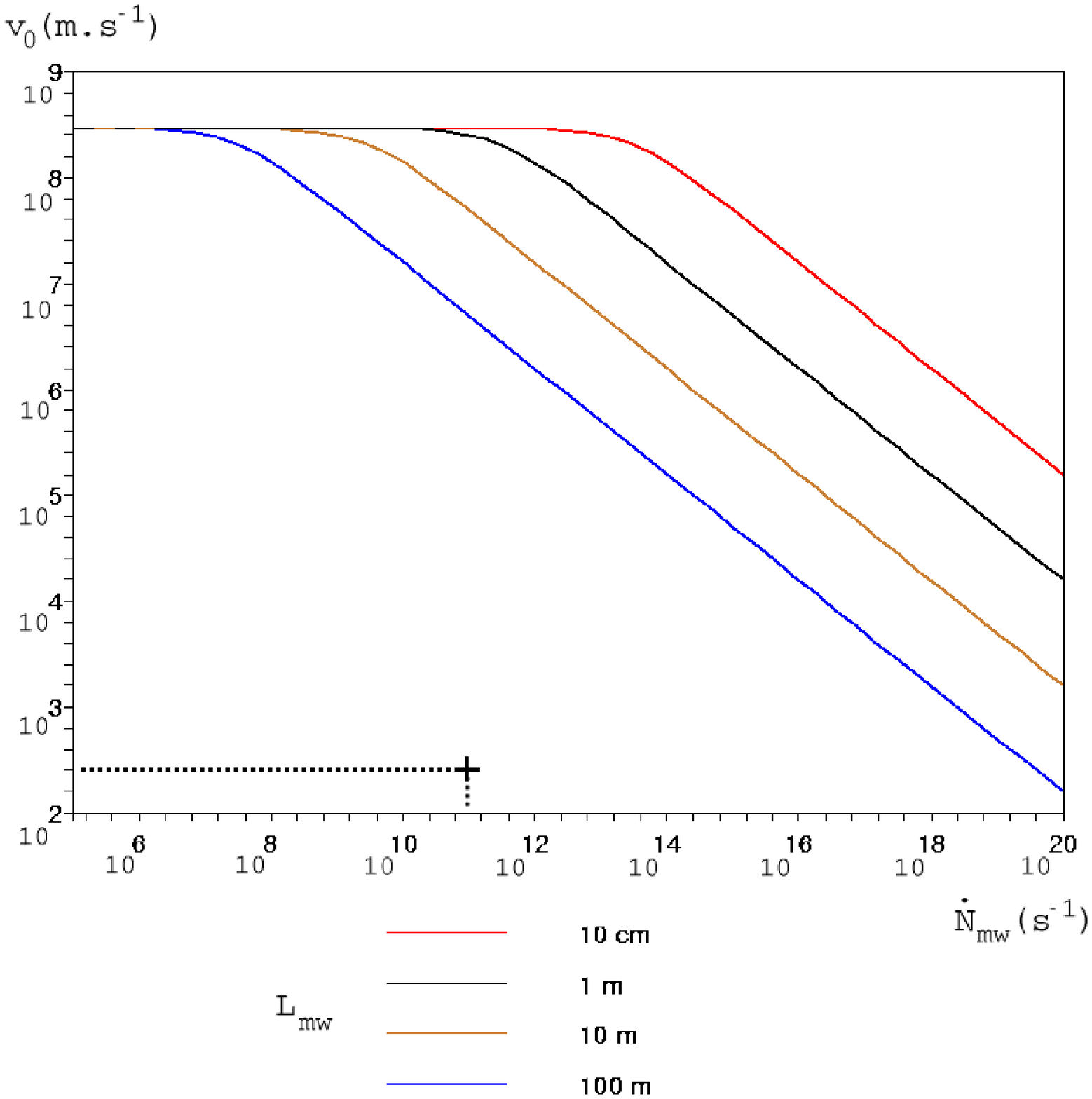}
      \caption{Required caracteristics of a MWI necessary to reach the sensitivity of Virgo.}
      \label{GRAPH1}
   \end{minipage} \hfill
   \begin{minipage}[l]{.47\linewidth}
      \includegraphics[width=\linewidth]{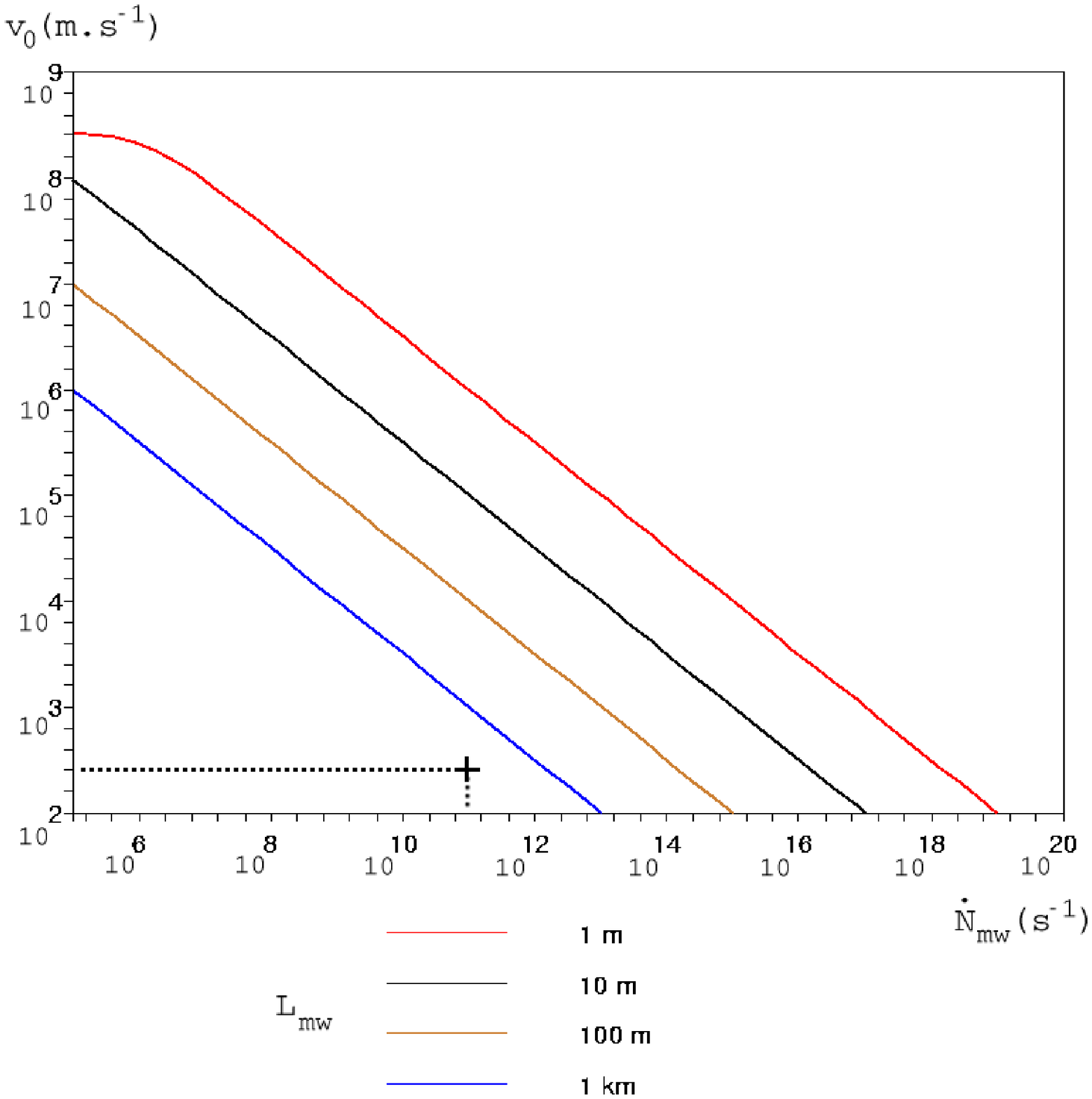}
      \caption{Required caracteristics of a MWI necessary to reach the sensitivity of LISA.}
      \label{GRAPH2}
   \end{minipage}
\end{figure}

On Fig.~\ref{GRAPH1} and \ref{GRAPH2} we represent the required characteristics\footnote{These curves are drawn from formula (\ref{COMP}) for the caesium mass. $v_0$, $L_\mathrm{mw}$ and $\dot{N}_\mathrm{mw}$ are respectively the initial atom wave group velocity, the MWI arm lenght and the atom flux. The characteristic point corresponds to the MWI described in \cite{gustavson00}.} of a MWI necessary to reach the sensitivity of Virgo (Fig.~\ref{GRAPH1}) and that of LISA (Fig.~\ref{GRAPH2}). The characteristic point of the MWI described in \cite{gustavson00} has been represented on both figures.

It appears clearly that the atom velocity has to be very high in order to reach Virgo sensitivity. However it would be a difficult challenge to keep the coherence and to separate the matter ray at such velocities. An idea could be to accelerate atoms inside the interferometer arms after separating the matter ray, and decelerate it before the reflections. For example, in GANIL\footnote{Grand Accelerateur National d'Ions Lourds}, one can obtain an ion ray at half the light velocity with fluxes up to 2,6.10$^{13}$~s$^{-1}$. In order to decrease the velocity a possibility could be to increase the matter
flux or the mass of the atoms \cite{arndt05} or to increase the arm length. However a good compromise cannot be achieved nowadays.

On the other hand, it seems easier to reach LISA sensitivity since the required atom velocity is much lower. The characteristic point of the MWI described in \cite{gustavson00} corresponds to a kilometric interferometer. A one meter MWI with $\sim 1,000$ round-trips in each arm would play the same role. It would be similar to the Fabry-Perot cavities in Virgo like interferometers.

\section{The rigid interferometer}

\subsection{The Fermi Coordinates (FC)}

In usual matter wave interferometers the mirrors and the beam splitters are "rigidly" bounded \cite{keith91,gustavson00,rasel95}. The concept of rigid body is not a relativistic one, however if the experimental set-up is much smaller than the gravitational wavelength ($L \ll \Lambda )$, one can introduce Fermi like coordinates (FC) comoving with the matter of the set-up supposed to be "rigid" \cite[pp.~39-47]{barone00}. We will assume here that a rigid body has fixed coordinates in the FC.

Lasers are used as mirrors and beam splitters in the experiments \cite{gustavson00,rasel95}~: it does not add a major perturbation to the phase difference. Indeed, in FC, the description of the physical phenomena (including the Maxwell equations) is very similar to the special relativistic one \cite[p.~52]{barone00}.

We choose the center of mass of the interferometer as the origin of the Fermi reference frame. The whole device is supposed to be free of constraints, so that the worldline of its center of mass is a geodesic. The metric can be derived from the general formula of Manasse \& Misner~\cite{manasse63}. It is convenient here to derive the coordinate transformation from EC, $x^\alpha$, to FC, $X^\alpha$, in order to link the movement of the optical elements in the two coordinate systems. This coordinate transformation has been derived with a general method that will be discussed in detail in a future paper. In order to find the metric up to the second order relatively to the $X^i$, one needs the coordinate transformation up to the third order~:

\begin{eqnarray}
\left\{ 
\begin{array}{l}
x^{r} = X^{r} - \frac{1}{2} h_{s}^{r} X^{s} - \frac{1}{3} \ddot{h}_{s}^{r} X^{s} Z^{2} + O (\xi^{4},h^{2}) \\[0.2cm] 
x^{a} = X^{a} - \frac{1}{4} \dot{h}_{rs} X^{r} X^{s} - \frac{1}{6} \ddot{h}_{rs} X^{r} X^{s} Z + O (\xi^{4},h^{2})
\end{array}
\right.
\end{eqnarray}

where $r,s=1$ or 2, $a=0$ or 3, $Z \equiv X^{3}$, $\xi \sim \sup{|X^i|}$ and $c=1$ (geometrical units). Then, the metric is~:

\begin{eqnarray}
\label{FCM}
\d s^2 & = & \eta _{\mu \nu} \d X^\mu \d X^\nu + \frac{1}{2} \ddot{h}_{rs} X^r X^s \left( \d T^2 + \frac{1}{3} \d Z^2 \right) - \frac{2}{3} \ddot{h}_{rs} X^r X^s \d T \d Z \nonumber \\
     &   & + \frac{1}{3} \ddot{h}_{rs} Z \d X^r \left( 2 X^s \d T - X^s \d Z + \frac{1}{2} Z \d X^s \right) + O(\xi^3,h^2)
\end{eqnarray}

where $T \equiv X^{0}$. This metric is in agreement with reference~\cite{fortini82} where Fortini \& Gualdi obtain the metric up to any order, although they do not derive the coordinate transformation that we consider here.

In the neighborhood of the plane $Z=0$, one obtains~:

\begin{equation}
\label{ASH}
\left\{ 
\begin{array}{l}
x^{r} = X^{r} - \frac{1}{2} h_{s}^{r} X^{s} + O (\xi^{4},h^{2}) \\[0.2cm] 
x^{a} = X^{a} - \frac{1}{4} \dot{h}_{rs} X^{r} X^{s} + O (\xi^{4},h^{2})
\end{array}
\right.
\end{equation}

and~:

\begin{equation}
\label{FC2}
\d s^2 = \eta_{\mu \nu} \d X^\mu \d X^\nu + \frac{1}{2} \ddot{h}_{rs} X^r X^s \d T^2 + O(\xi^3,h^2)
\end{equation}

The coordinate transformation (\ref{ASH}) was first found by Ashby \& Dreitlein in reference~\cite{ashby75}. However, contrary to the claim of the authors, the metric~(\ref{FC2}) (that can be derived from~(\ref{ASH})) is a Fermi metric in the sense of Manasse \& Misner \cite{manasse63} only in the plane $Z = 0$, and not in the whole space.

Now that we have the metric tensor in the FC, we will compute the phase difference for several rigid interferometer.

\subsection{The Michelson-Morley configuration}

First we study the Michelson-Morley configuration of Fig.~\ref{MICHPIC}, but now we assume that the coordinates of the optical elements are constant in the FC. From eqs.~(\ref{PHITOT}) and (\ref{FC2}) we find~:

\begin{equation}
\label{RIG}
\Delta \phi =4\pi h_{+}\cdot \frac{L}{\lambda }\cdot \left( 1-\frac{V}{\Omega L}\sin {\frac{\Omega L}{V}}\right) \cdot \sin {\Omega t}
\end{equation}

with the notations of section~\ref{PD}. This formula is in agreement with the one obtained in \cite{roura04}. The assumption $L\ll \Lambda $ implies $\Omega L/c\ll 2\pi $. Therefore, for a LWI, formula~(\ref{RIG}) reduces to~:

\begin{equation}
\label{NULL}
\Delta \phi_\mathrm{lw} \simeq 0
\end{equation}

This result is well-known~: if the arms of a LWI were rigid there would be no signal. The situation is different for a MWI where we can consider two different regimes~\cite{roura04}~:

\begin{itemize}

\item $L \ll \dfrac{v_{0}}{c} \cdot \Lambda$ (\ie{} $\Omega L / V \ll 2 \pi$)~: formula (\ref{RIG}) reduces to $\Delta \phi _\mathrm{mw}\simeq 0$. This regime occurs when the flight time of an atom in the interferometer is much less than the GW period. In this case there is no signal.

\item $\dfrac{v_{0}}{c} \cdot \Lambda \ll L$ (\ie{} $\Omega L / V \gg 2 \pi$)~: one can assume $V / \Omega L \ll 1.$ This regime occurs when the flight time is much longer than the period of the GW. In this case the amplitude of the phase difference in (\ref{RIG}) reduces to~:

\begin{equation}
\label{DIFF2}
\widetilde{\Delta \phi }_\mathrm{mw}\simeq 4\pi h_{+}\cdot \frac{L}{\lambda}
\end{equation}

This new specific regime has no equivalent with free interferometers nor with a rigid LWI. One can notice that the present amplitude given by (\ref{DIFF2}) is similar to formula (\ref{DIFF}). Unfortunately in the present regime the sensitivity is limited by the condition $v_{0}\ll \Omega L.$

\end{itemize}

\subsection{The Ramsey-Bord\'{e} configuration}

\label{SECBLO}

\begin{figure}[h]
\begin{center}
\includegraphics[width=0.5\linewidth]{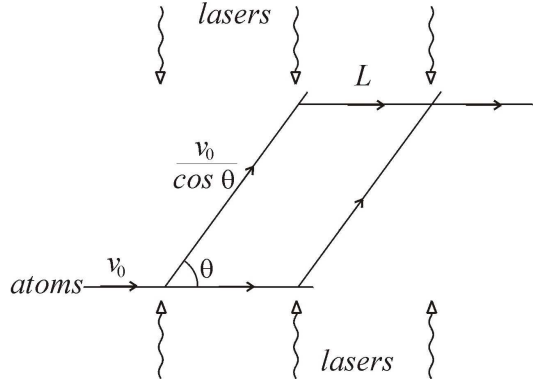}
\caption{\footnotesize Ramsey-Bord\'{e} interferometer in the $X$-$Y$ plane.}
\label{BORDE}
\end{center}
\end{figure}

We study now a MWI with a Ramsey-Bord\'{e} configuration \cite{keith91,gustavson00,rasel95} represented on Fig.~\ref{BORDE}. We define $X=X^1$ and $Y=X^2$. Atoms move along the $X$ axis and we assume that $v_0 \ll c$. At the output of the interferometer, we calculate from eqs.~(\ref{PHITOT}) and (\ref{FC2}) the phase difference between the two beams~:

\begin{equation}
\label{RAM}
\Delta \phi =\Delta \phi _{+}+\Delta \phi _{\times }
\end{equation}

with~:

\begin{equation}
\Delta \phi_{+} = - 4 \pi h_{+} \dfrac{L}{\lambda} \sin{\psi} \tan^2{\theta} \left[ \cos {\left( \Omega t+\psi \right) }+\dfrac{\sin {\psi }}{2\psi } \cos {\left( \Omega t\right) }\right]
\end{equation}

\begin{equation}
\Delta \phi_{\times} = - 4 \pi h_{\times} \dfrac{L}{\lambda} \cos{\psi} \tan{\theta} \left[ \sin {\left( \Omega t+\varphi _{\times }-\psi \right) }-\dfrac{\sin {\psi }}{\psi }\tan {\psi }\cos {\left( \Omega t+\varphi _{\times }\right) }\right]
\end{equation}

where $\theta$ is the separation angle and $\psi =\dfrac{\Omega L}{2v_{0}}$.

The lasers give to the atoms a velocity $\Delta v = v_0 \tan{\theta}$ along the $Y$ axis. In order to achieve a high value of $\theta$, one must communicate to the atom a high momentum during the interactions with the photons. This is very difficult because this has to be done without any loss of coherence. The momentum of the photons being $\hbar k$, one can imagine to transfer $m \Delta v = N \times \hbar k$ to the atom. Significant results, $N \simeq 1000$, have already been obtained along this line of research \cite{Battesti04,clade_these}; they still remain insufficient today.

We assume now that $\theta \simeq \pi / 4$ so that $\tan{\theta} \simeq 1$. Here again two regimes can be considered~:

\begin{itemize}

\item $L\ll \dfrac{v_{0}}{c}\cdot \Lambda $ (ie) $\psi = \dfrac{\Omega L}{2 v_{0}} \ll 2\pi$~: in this case

\begin{equation}
\label{DIFFnew}
\Delta \phi \simeq -4 \pi h_{\times} \cdot \frac{L}{\lambda }\cdot \sin {\left(\Omega t+\varphi _{\times }\right) }
\end{equation}

This expression results in an amplitude already given, formula (\ref{DIFF}) above. Therefore, Figs.~\ref{GRAPH1} and~\ref{GRAPH2} can be used for the discussion of the experimental design of the interferometer (with $h_{+}\rightarrow h_{\times})$.

\item $\dfrac{v_{0}}{c}\cdot \Lambda \ll L$ (ie) $2 \pi \ll \dfrac{\Omega L}{2 v_0} = \psi$~: expression (\ref{RAM}) gives

\begin{equation}
\Delta \phi \simeq - 4 \pi \cdot \frac{L}{\lambda } \cdot \left[ h_{+}\sin {\psi } \cos {\left( \Omega t+\psi \right) }+h_{\times }\cos {\psi }\sin {\left(\Omega t+\varphi _{\times }-\psi \right) }\right]
\end{equation}

The phase difference $\Delta \phi $ is a periodic function of the time whose amplitude displays the same order of magnitude that was previously put forward in expressions~(\ref{DIFF}) and~(\ref{DIFFnew}). Here, however, one can choose the value of $\psi $ in order to measure $h_{+}$ or $h_{\times }$. This is a positive point but the sensitivity remains however limited by the low values of $v_{0}$ in this regime.

\end{itemize}

\section{Conclusion}

In this paper we considered matter wave and light wave interferometers designed to detect gravitational waves. This has been considered already in the literature in some special cases \cite{chiao04a,roura04,foffa06} where several different claims correspond to different assumptions (more or less justified) about the coordinate systems comoving with the interferometer.

We considered "free" and "rigid" interferometers. The comoving coordinates have been chosen as Einstein coordinates in the first case; it seems that this choice raises no discussion. The comoving coordinates have been chosen as Fermi-like coordinates in the second case. This choice is very natural because one can show that the first approximation of the equations of a continuous medium are ordinary non relativistic equations with just an extra gravitational force density (the mechanical detectors of gravitational waves are precisely based on such equations).

We have considered only plane interferometers orthogonal to the propagation of the gravitational waves. If we change the orientation of the interferometer relatively to the gravitational wave, the sensitivity is slightly modified but the orders of magnitude remain the same. Moreover we only considered periodic waves and no pulses of gravitational waves. We believe that the comparison between light-wave interferometers and matter-wave interferometers is not deeply affected by such a simplification.

In the cases that we studied we obtained an estimation of the sensitivity of matter-wave interferometers. We especially considered the shot noise limit, however, it is the thermal noise which most troublesome at the present moment. The answer of this problem could be the construction of a compact (one meter) interferometer which could be cooled at very low temperature. In order to estimate roughly the thermal noise we consider that the interferometer is fixed on a bench of mass $M\sim 500$~kg which displays an eigenfrequency of order $\omega _{0}\sim 10^4$~s$^{-1}$. Therefore,
following reference~\cite{punturo04}, one finds the limit~:

\begin{equation}
h_{min}\sim \frac{1}{L}\left( \frac{4k_{B}T}{MQ\omega _{0}^{2}\Omega } \right) ^{1/2}
\end{equation}

where $k_{B}$ is the Boltzmann constant. With a quality factor $Q\sim 10^{7}, $ and a temperature $T\sim 10^{-2}$ K one finds $h_{min} \sim 10^{-20}$ for $\Omega \sim 10^{-2}$~s$^{-1}$.

Now if a clear conclusion had to be taken from the previous estimations, we would claim that in the future, compact, very low temperature matter-wave interferometers will not be a serious challenger to high frequency detectors of gravitational waves (such as Virgo or LIGO) but to LISA. Of course major improvements still remain necessary today.




\begin{thebibliography}{10}
\expandafter\ifx\csname url\endcsname\relax
  \def\url#1{\texttt{#1}}\fi
\expandafter\ifx\csname urlprefix\endcsname\relax\def\urlprefix{URL }\fi

\bibitem{abbott05b}
{B. Abbott \emph{et al.} (LIGO Collaboration)}, {Upper limits from the LIGO and
  TAMA detectors on the rate of gravitational-wave bursts}, Phys. Rev. D
  72~(12) (2005) 122004--+.

\bibitem{abbott05a}
{B. Abbott \emph{et al.} (LIGO Collaboration)}, {Upper Limits on a Stochastic
  Background of Gravitational Waves}, Phys. Rev. Lett. 95~(22) (2005)
  221101--+.

\bibitem{abbott04}
{B. Abbott \emph{et al.} (LIGO Collaboration)}, {Setting upper limits on the
  strength of periodic gravitational waves from PSR J1939+2134 using the first
  science data from the GEO 600 and LIGO detectors}, Phys. Rev. D 69~(8) (2004)
  082004--+.

\bibitem{lisa00}
ESA-SCI, {LISA: A Cornerstone Mission for the Observation of Gravitationnal
  Waves}, Tech. rep., ESA-SCI,
  {ftp://ftp.rzg.mpg.de/pub/grav/lisa/sts/sts\_1.05.pdf} (2000).

\bibitem{chiao04a}
R.~Y. {Chiao}, A.~D. {Speliotopoulos}, {Towards MIGO, the matter-wave
  interferometric gravitational-wave observatory, and the intersection of
  quantum mechanics with general relativity}, J. Mod. Opt. 51 (2004) 861--899.

\bibitem{roura04}
A.~{Roura}, D.~R. {Brill}, B.~L. {Hu}, C.~W. {Misner}, W.~D. {Phillips},
  {Gravitational wave detectors based on matter wave interferometers (MIGO) are
  no better than laser interferometers (LIGO)}, ArXiv General Relativity and
  Quantum Cosmology e-prints.

\bibitem{foffa06}
S.~{Foffa}, A.~{Gasparini}, M.~{Papucci}, R.~{Sturani}, {Sensitivity of a small
  matter-wave interferometer to gravitational waves}, Phys. Rev. D 73~(2)
  (2006) 022001--+.

\bibitem{linet76}
B.~{Linet}, P.~{Tourrenc}, {Changement de phase dans un champ de gravitation:
  Possibilit{\'e} de d{\'e}tection interf{\'e}rentielle}, Can. J. Phys. 54
  (1976) 1129.

\bibitem{stodolsky79}
L.~{Stodolsky}, {Matter and light wave interferometry in gravitational fields},
  Gen. Relativ. Gravitation 11 (1979) 391--405.

\bibitem{barone00}
M.~{Barone}, G.~{Calamai}, M.~{Mazzoni}, R.~{Stanga}, F.~{Vetrano} (Eds.),
  {Experimental Physics of Gravitational Waves}, 2000.

\bibitem{keith91}
D.~W. {Keith}, C.~R. {Ekstrom}, Q.~A. {Turchette}, D.~E. {Pritchard}, {An
  interferometer for atoms}, Phys. Rev. Lett. 66 (1991) 2693--2696.

\bibitem{gustavson00}
T.~L. {Gustavson}, A.~{Landragin}, M.~A. {Kasevich}, {Rotation sensing with a
  dual atom-interferometer Sagnac gyroscope }, Classical Quantum Gravity 17
  (2000) 2385--2398.

\bibitem{punturo04}
M.~{Punturo}, The virgo sensitivity curve, Virgo note, code:
  VIR-NOT-PER-1390-51.

\bibitem{arndt05}
M.~{Arndt}, L.~{Hackerm{\"u}ller}, E.~{Reiger}, {Interferometry with Large
  Molecules: Exploration of Coherence, Decoherence and Novel Beam Methods},
  Braz. J. Phys. 35 (2005) 216--223.

\bibitem{rasel95}
E.~M. {Rasel}, M.~K. {Oberthaler}, H.~{Batelaan}, J.~{Schmiedmayer},
  A.~{Zeilinger}, {Atom Wave Interferometry with Diffraction Gratings of
  Light}, Phys. Rev. Lett. 75 (1995) 2633--2637.

\bibitem{manasse63}
F.~K. Manasse, C.~W. Misner, {Fermi Normal Coordinates and Some Basic Concepts
  in Differential Geometry}, J. Math. Phys. 4~(6) (1963) 735--745.
\newline\urlprefix\url{http://link.aip.org/link/?JMP/4/735/1}

\bibitem{fortini82}
P.~L. {Fortini}, C.~{Gualdi}, {Fermi normal co-ordinate system and
  electromagnetic detectors of gravitational waves. I - Calculation of the
  metric}, Nuovo Cimento B 71 (1982) 37--54.

\bibitem{ashby75}
N.~{Ashby}, J.~{Dreitlein}, {Gravitational wave reception by a sphere}, Phys.
  Rev. D 12 (1975) 336--349.

\bibitem{Battesti04}
R.~{Battesti}, P.~{Clad{\'e}}, S.~{Guellati-Kh{\'e}lifa}, C.~{Schwob},
  B.~{Gr{\'e}maud}, F.~{Nez}, L.~{Julien}, F.~{Biraben}, {Bloch Oscillations of
  Ultracold Atoms: A Tool for a Metrological Determination of $h/m_{Rb}$},
  Phys. Rev. Lett. 92~(25) (2004) 253001.

\bibitem{clade_these}
P.~{Clad{\'e}}, {Oscillations de Bloch d'atomes ultrafroids et mesures de la
  constante de structure fine}, Ph.D. thesis, {Universit\'e Pierre et Marie
  Curie} (Oct. 2005).
\newline\urlprefix\url{http://tel.ccsd.cnrs.fr/documents/archives0/00/01/07/30%
/index\_fr.html}

\end{thebibliography}
\end{document}